# Simulation of Two-qubit Gate Variability and Fidelity of Spin Qubits Built on Nanosheet Technology

Trung Nguyen
Electrical Engineering
San Jose State University
San Jose, USA
trung.nguyen03@sjsu.edu

Sarah Dweik
Electrical Engineering
San Jose State University
San Jose, USA
sarah.dweik@sjsu.edu

Hiu Yung Wong
Electrical Engineering
San Jose State University
San Jose, USA
hiuyung.wong@sjsu.edu*

***Abstract*—Silicon spin qubits are promising for large-scale quantum-computer integration because they can fully leverage the well-developed semiconductor infrastructure. However, the low fidelity of two-qubit entanglement gates remains a key barrier to large-scale integrations. Recent simulations of silicon spin-qubit two-qubit gates have been performed on silicon-on-insulator (SOI) platforms, while nanosheet-based charge-qubit work has been limited to single-qubit operation using a two-dimensional Schrödinger approximation. In this work, we study silicon spin-qubit double quantum dots built on nanosheet technology using the Quantum Technology Computer-Aided Design (QTCAD) simulation suite to run three-dimensional Poisson and Schrödinger solvers, followed by a many-body solver to extract exchange interactions. We evaluate the exchange energy sensitivity to process and bias variations and then use QuTiP to solve the master equation for a two-qubit gate. The results show that millivolt-level bias variations at the plunger and middle barrier gates can reduce the gate fidelity below 99%, a common threshold target for many fault-tolerant quantum-computing algorithms. Gate-referred 1/f charge-noise effects are also analyzed through the resulting coherence time.**

## I. INTRODUCTION

Single-electron spin qubits in semiconductor quantum dots have each qubit encoded in the spin state of one confined electron [1][2][3]. Single-qubit operations are usually performed using spin-resonance techniques such as electron spin resonance (ESR) or electric-dipole spin resonance (EDSR), while two-qubit gates are realized by electrically controlling the exchange interaction between neighboring electron spins [4] or other methods.

Exchange-based two-qubit gates for single electron spin qubit have been experimentally demonstrated in silicon MOS and Si/SiGe double quantum dots (DQDs) [5][6][7][8], with recent demonstrations achieving gate fidelities exceeding 99%. However, maintaining high fidelity under charge noise, and process variability remains a key challenge. Simulation studies have therefore investigated how two-qubit gate operation is affected by charge noise, device electrostatics, and charge defects [9][10][11]. In particular, a device-to-gate-fidelity modeling framework for semiconductor DQDs was developed in [10], including Si-MOS examples, and recent simulation work on fully depleted silicon-on-insulator (FD-SOI) DQDs [11] used advanced Quantum Technology Computer Aided Design (QTCAD) software suite to evaluate

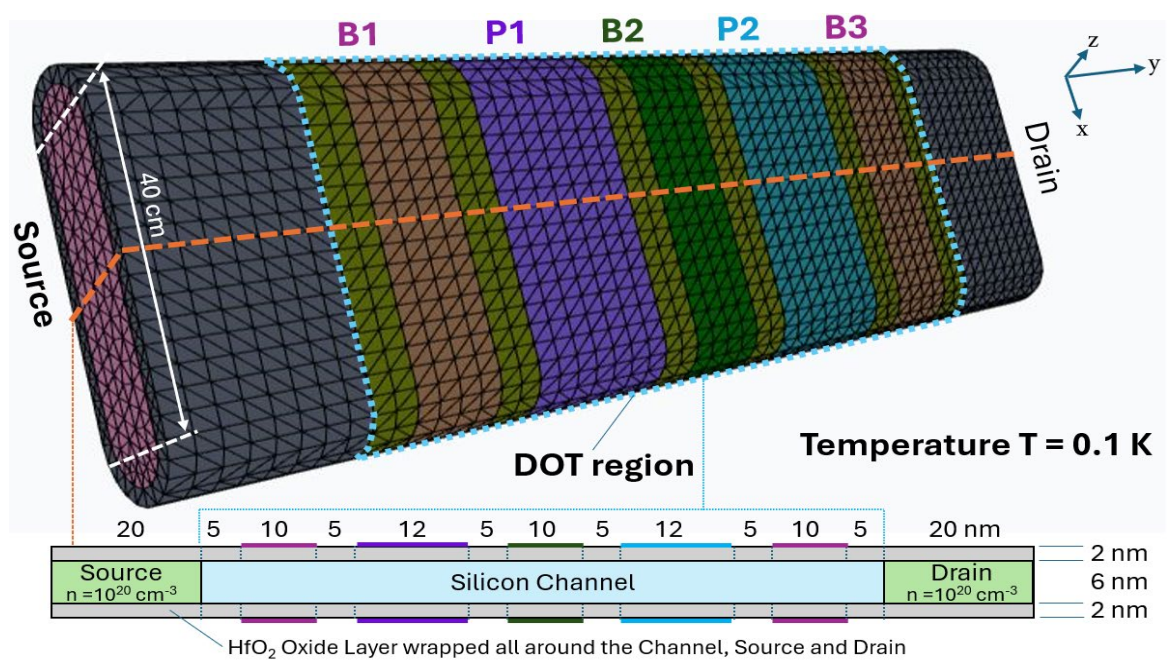

Fig. 1: Top: 3D view of the DQD nanosheet device with a silicon channel, n-doped source/drain regions, $HfO_2$ gate oxide, plunger gates P1 and P2, and barrier gates B1, B2, and B3. Bottom: y-z cross-section at the dotted region.

two-qubit gate fidelity under charge-defect-induced exchange fluctuations.

Nanosheet (NS) technology offers strong gate electrostatics and is closely related to advanced nanosheet gate-all-around (GAA) transistor platforms, which have been introduced into leading-edge CMOS manufacturing [12]. These features make NS devices attractive for gate-defined quantum dots, but device-level studies connecting nanosheet geometry, exchange variability, and two-qubit gate fidelity remain largely unexplored. A previous nanosheet qubit simulation work [13] studied a charge-qubit implementation, but it was limited to a single-qubit gate and used a two-dimensional Schrödinger approximation.

In this paper, we model a nanosheet DQD spin-qubit device using QTCAD [14]. The simulation workflow combines three-dimensional nonlinear Poisson electrostatics, single-electron Schrödinger calculations, and many-body configuration-interaction treatment of electron-electron interactions. The extracted exchange energy is then used to study two-qubit gate fidelity under bias variations, process variations, and gate-referred 1/f charge noise. The main contributions of this work include: (i) the design of a nanosheet spin-qubit DQD, (ii) the extraction of exchange-interaction variability across relevant bias and geometric parameters, and (iii) a gate-level fidelity and decoherence analysis driven by the simulated exchange interaction.

## II. TWO-QUBIT GATE FOR SILICON SPIN QUBITS AND SIMULATION WORKFLOW

A two-qubit entanglement gate in a silicon electron spin-qubit system is best understood by the interaction between the singlet ($S$) and triplet states ($T_+, T_0, T_-$, which are degenerate when no magnetic field is applied) of two electrons, with one electron confined in each dot of the double-quantum-dot (DQD) (Figs. 1 and 2).

*Corresponding author: hiuyung.wong@sjsu.edu

The exchange interaction strength $J = E(T_0) - E(S)$ is the energy gap between the triplet and singlet states and determines how the two qubits evolve and interact with each other through the exchange Hamiltonian, $H(t) = J(t)S_1 \cdot S_2$, where $S_1, S_2$ are the spin angular momentum operator in each dot of DQD [3]. Electrically pulsing the interdot barrier changes the wavefunction overlap between the two dots and therefore controls $J(t)$.

The time evolution generated by this Hamiltonian is determined by the accumulated exchange phase $\theta = \frac{1}{\hbar}\int_0^{t_g} J(t)dt$, which determines the two-qubit operation. For example, $\theta = \pi$ gives a $SWAP$ operation, while $\theta = \pi/2$ gives a $\sqrt{SWAP}$ operation, which is an entangling two-qubit gate. Together with single-qubit rotations, such exchange pulses can be used to construct universal two-qubit logic gates such as CNOT or controlled-phase gates.

QTCAD is used for the simulation. To perform an accurate simulation, the following procedure was followed:

1. The nanosheet device is first defined in QTCAD on a high-density finite-element mesh with approximately $5 \times 10^5$ nodes with a finer mesh size of 0.25-0.5nm in the dot region, is imported into QTCAD.
2. A three-dimensional nonlinear Poisson solver is used to compute the electrostatic confinement potential across the device, with the quantum dot region specified, setting the charge density $\rho[\phi]$ to zero.
3. An effective detuning offset is extracted by sweeping the relative plunger-gate bias to compensate for mesh-induced asymmetries. The offset is chosen at the most symmetric DQD bias point, identified by the minimum tunnel coupling between the two localized dot states. The offset is then added to the plunger gate voltages in the subsequent simulations. The offset depends on the geometry and bias and it ranges from 0.5 μV to 0.2 mV in this paper.
4. The resulting electrostatic potential is used in a single-electron Schrödinger solver to obtain the energy eigenvalues and corresponding wavefunctions.
5. These single-electron states are passed to a many-body solver, where wavefunctions integrals are evaluated to construct the electron-electron Coulomb interaction matrix elements for the two, three, and four-electron confinement configurations.
6. The resulting many-body Hamiltonians are diagonalized, and the exchange interaction is extracted from the singlet-triplet energy splitting in the two-electron confinement regime, $J = E_T - E_S$.
7. The average number of confined electrons is evaluated from the many-body subspaces to verify that the system remains in the desired two-electron confinement regime.
8. For convergence, a tolerance level of $10^{-10}$ is used for the Poisson and Schrödinger solvers; sixteen spatial orbitals are used in the Schrödinger solver, and twelve spatial orbitals are used in the exchange-interaction calculation.

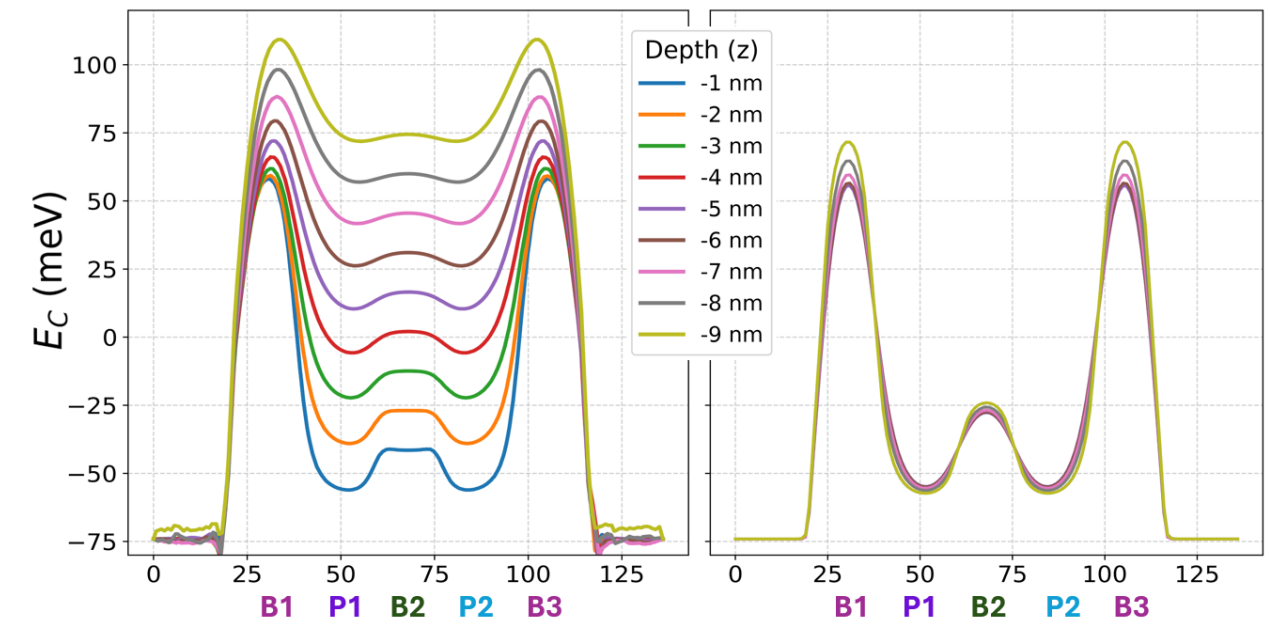

Fig. 2: Conduction band edge profiles at various z-locations along y (Fig. 1) of FD-SOI (left) and NS (right) of the same layout and silicon thickness.

In this simulation workflow, a sequential Poisson–Schrödinger solver is used, in which the electrostatic confinement potential is first obtained from the Poisson solver and then used as input to the single-particle Schrödinger equation; the Schrödinger electron density is not iteratively fed back into the Poisson solution. This approximation is reasonable for the two-electron DQD regime considered here, where the confinement potential is mainly set by the gate biases, geometry, dielectric environment, and source/drain reservoirs. The electron-electron Coulomb interaction is still included in the many-body calculation used to extract the singlet and triplet energies. Therefore, neglecting the self-consistent Schrödinger–Poisson loop is expected to have a small effect on the qualitative trends in exchange interaction and gate fidelity.

## III. Simulation Structure and Results

### a) Nanosheet Design

The simulated nanosheet device consists of an undoped silicon channel with n-doped source and drain region at $10^{20}$ $cm^{-3}$ and a 2 nm $HfO_2$ gate dielectric (Fig. 1). It has three barrier gates (B1, B2, B3) and two plunger gates (P1, P2) to form the DQD. The plunger gates control the depths of the two quantum wells, while the middle barrier gate B2 controls the inter-dot tunnel barrier, and therefore the exchange interaction. Compared to FD-SOI [11], NS has a nearly constant potential profile in the $z$-direction for the same layout and Si thickness (Fig. 2). As a result, different operating biases are required to obtain the target two-electron confinement condition. A nominal device is designed to have $V_{P1} = V_{P2} = V_{P12} = 0.595$V, $V_{B1} = V_{B3} = V_{B13} = 0.465$V, with $t_{Si} = 6$ nm, a plunger gate length $L_P = 12$ nm, and a barrier $L_B$ = 10nm. *Note that the offsets to eliminate the asymmetry due mesh is superimposed to the voltages mentioned.* Fig. 3 shows the electron density of $S$ and $T_0$. The extracted singlet and triplet spatial electron densities agree with the expected symmetry of the two-electron wavefunctions. The singlet state is spatially symmetric with more spatial overlap, while the triplet state is spatially anti-symmetric with less spatial overlap, validating that the simulated DQD supports the intended spin-qubit operating regime.

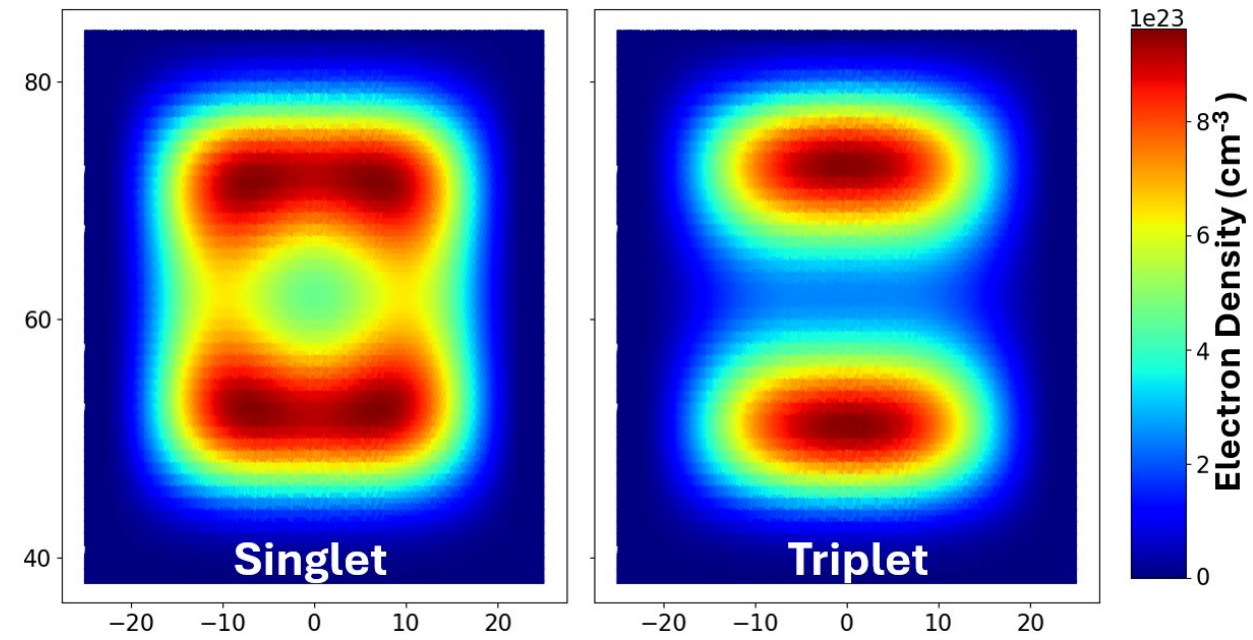

Fig. 3: DQD spatial electron density of the singlet and triplet states on the (x, y, 0) plane with $V_{B2} = 0.59$V.

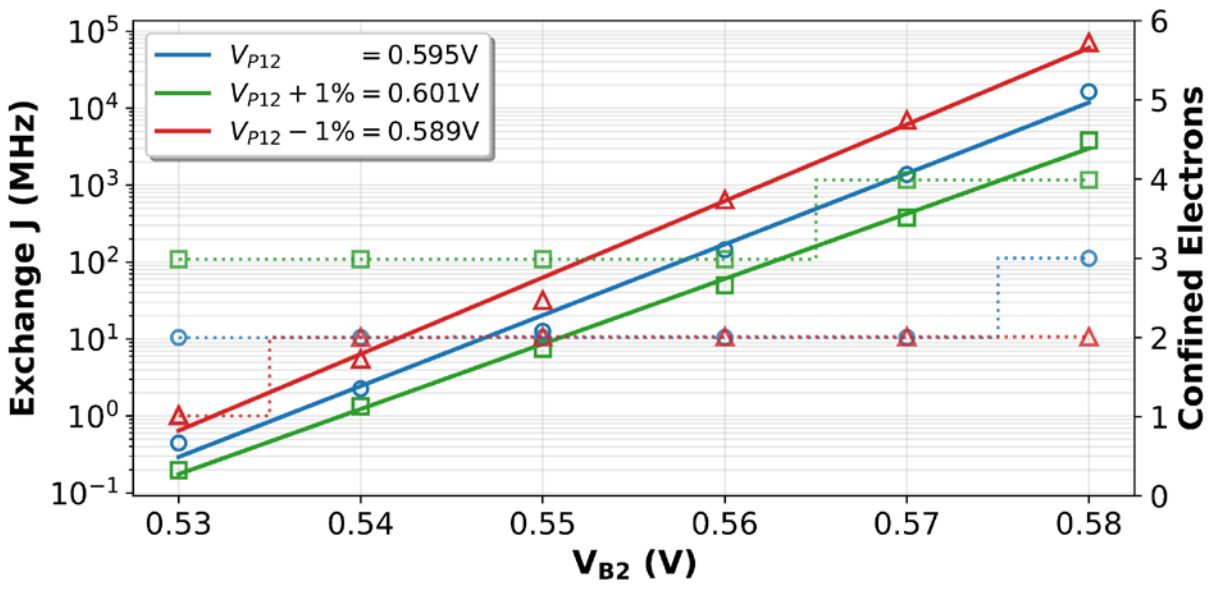

Fig. 4. DQD exchange interaction J and confined electron number as a function of $V_{B2}$ for $V_{P12}$ = 0.595V ± 1%.

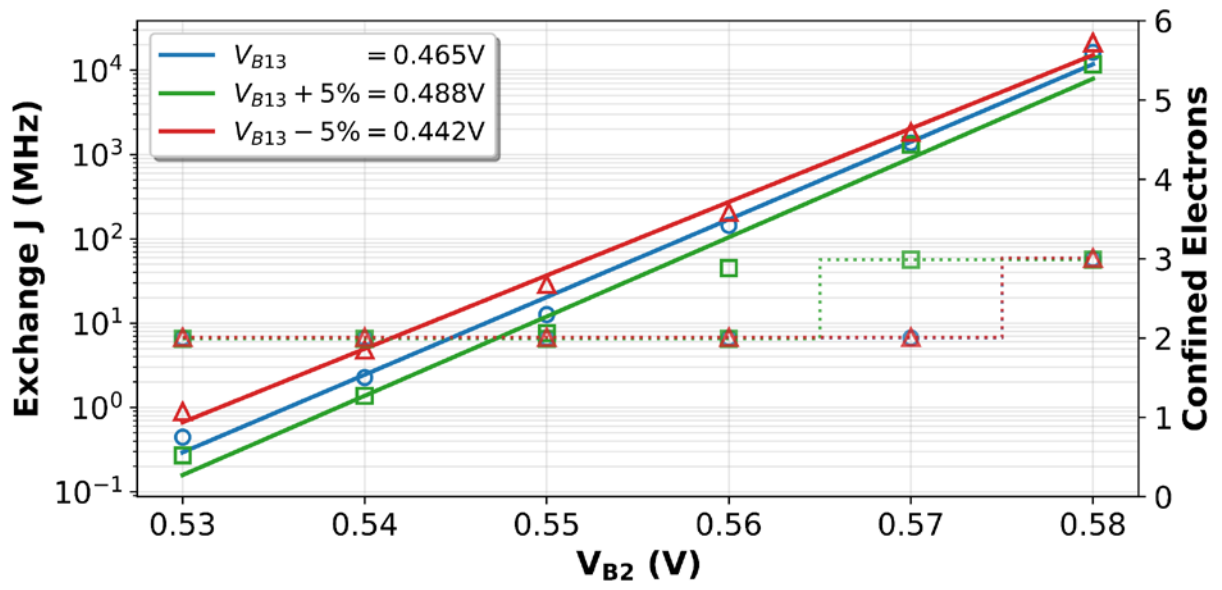

Fig. 5. DQD exchange interaction J and confined electron number as a function of $V_{B2}$ for $V_{B13}$ = 0.465V ± 5%

*b) Bias Variations*

The exchange interaction and confined electron number are first studied as a function of the plunger-gate voltage $V_{P12}$ and the middle-barrier voltage $V_{B2}$. For a representative operating point of $V_{B2}$ = 0.55 V, changing $V_{P12}$ by ±1% modifies J by approximately 4.2 times, while changing $V_{B2}$ by ±1% modifies J by approximately 6.2 times (Fig. 4). In contrast, changing the side barrier bias $V_{B13}$ by ±5% changes J by only ~3.8 times (Fig. 5). These results indicate that it is most critical to have accurate control and avoid background charges in the quantum well and the middle barrier regions. This is because the difference $V_{P12}$ - $V_{B2}$ sets the inter-dot barrier height, which strongly controls tunneling and therefore the exchange. In addition, only a limited bias window confines exactly two electrons, so both exchange control and electron occupation must be considered simultaneously.

*c) Geometry Variations*

Figs. 6, 7, and 8 plot the effects of variations in nanosheet thickness, $t_{Si}$, barrier gate length, $L_B$, and plunger gate length, $L_P$, on electron confinement and exchange energy.

A $t_{Si}$ variation of ±10% has a comparatively small impact on the exchange interaction but produces a larger impact on the number of confined electrons. This occurs because changing the nanosheet thickness directly modifies the vertical confinement energy and subband spacing. These effects shift the dot electrochemical potentials and therefore strongly affect electron occupation.

By contrast, $L_B$, variation has a significant impact on $J$ while having little effect on confined electron number. Since the exchange interaction depends strongly on tunneling through the middle barrier, variations in $L_B$ directly modify the tunnel coupling. The confined electron number is instead determined more strongly by the well depth and well width.

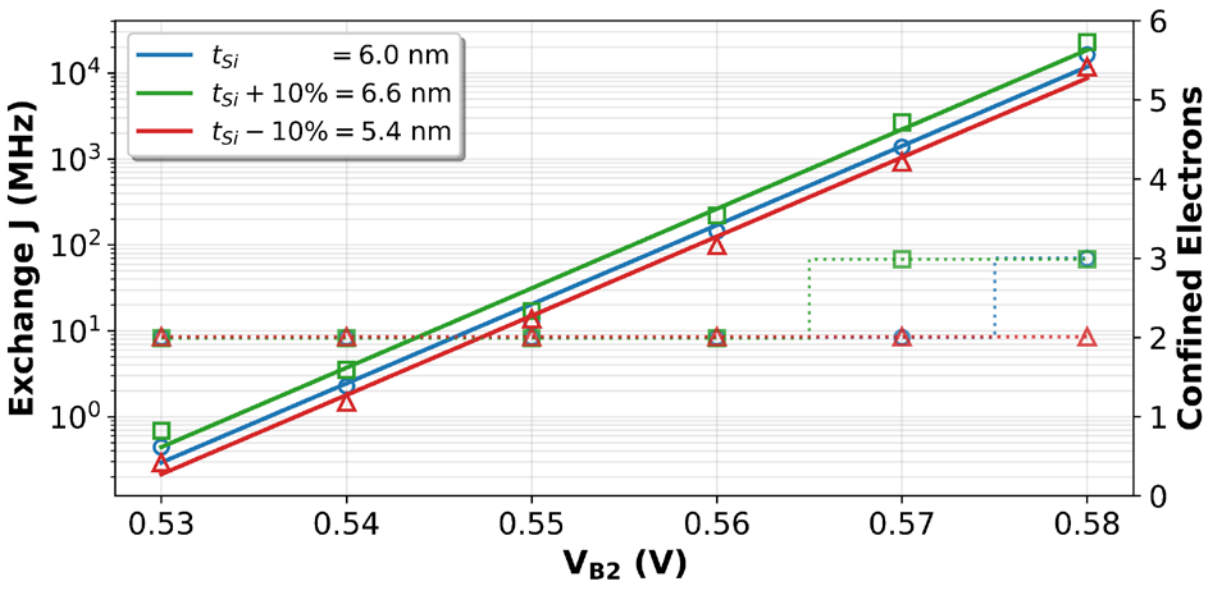

Fig. 6. DQD exchange interaction J and confined electron number as a function of $V_{B2}$ for $t_{Si}$ = 6.00 nm ± 10%

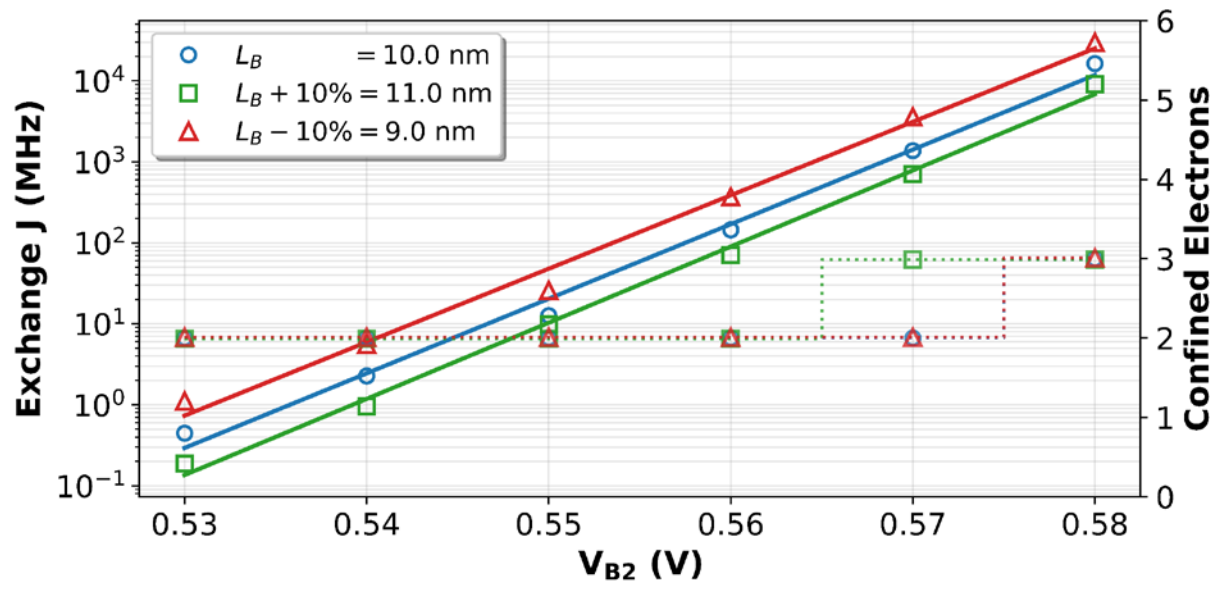

Fig. 7. DQD exchange interaction J and confined electron number as a function of $V_{B2}$ for $L_B$ = 10.00 nm ± 10%.

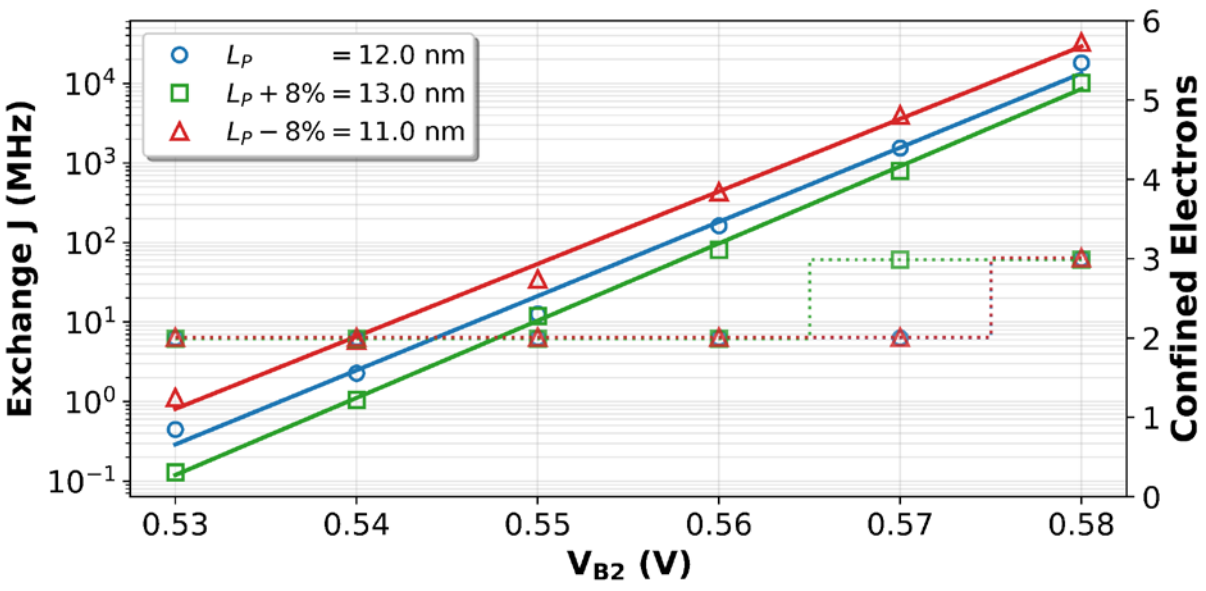

Fig. 8. DQD exchange interaction J and confined electron number as a function of $V_{B2}$ for $L_P$ = 12.00 nm ± 8%.

Plunger length $L_P$ variation affects both electron occupation and exchange interaction. By changing the effective lateral width of the dot potential well, $L_P$ modifies the orbital eigenenergies, which shifts the dot electrochemical potentials and influences the number of confined electrons. At the same time, changes in $L_P$ alter wavefunction localization and interdot overlap, thereby affecting the exchange interaction $J$.

*d) Fidelity and Decoherence Time*

Using the exchange interaction $J$ extracted from QTCAD, QuTip [15] is used to simulate the fidelity of a two-qubit gate under bias variations, and the exchange Hamiltonian $H = 2\pi J/4(\sigma_{x1}\sigma_{x2} + \sigma_{y1}\sigma_{y2} + \sigma_{z1}\sigma_{z2})$ was solved via mesolve. Using the nominal $J$ at $V_{B2}$ = 0.54 V, we find the gate time of the $\sqrt{SWAP}$ gate to be about 114.6 ns (Fig. 9). Bias variations result in $J$ variations and, thus, they have different fidelities if the gate time is still assumed to be 114.6 ns. The fidelity is calculated by using state fidelity $F(\psi, \boldsymbol{\sigma}) = \sqrt{\langle\psi|\boldsymbol{\sigma}|\psi\rangle}$, where $\boldsymbol{\sigma}$ is the density matrix of the state after a perfect $\sqrt{SWAP}$ gate and $|\psi\rangle$ is the 2-qubit state at time $t$. Bias variations of $V_{P12}$ ±1 mV, $V_{B2}$ ±1 mV, and $V_{B13}$ ±1 mV are studied. Fig. 9 shows the state evolution under nominal $J$ (top) and when $V_{B2}$ = 0.541 V (bottom), as an example. Fig. 10 shows that the fidelity can be less than 99% (threshold for many fault-tolerant quantum computing) for $V_{B2}$+1 mV.

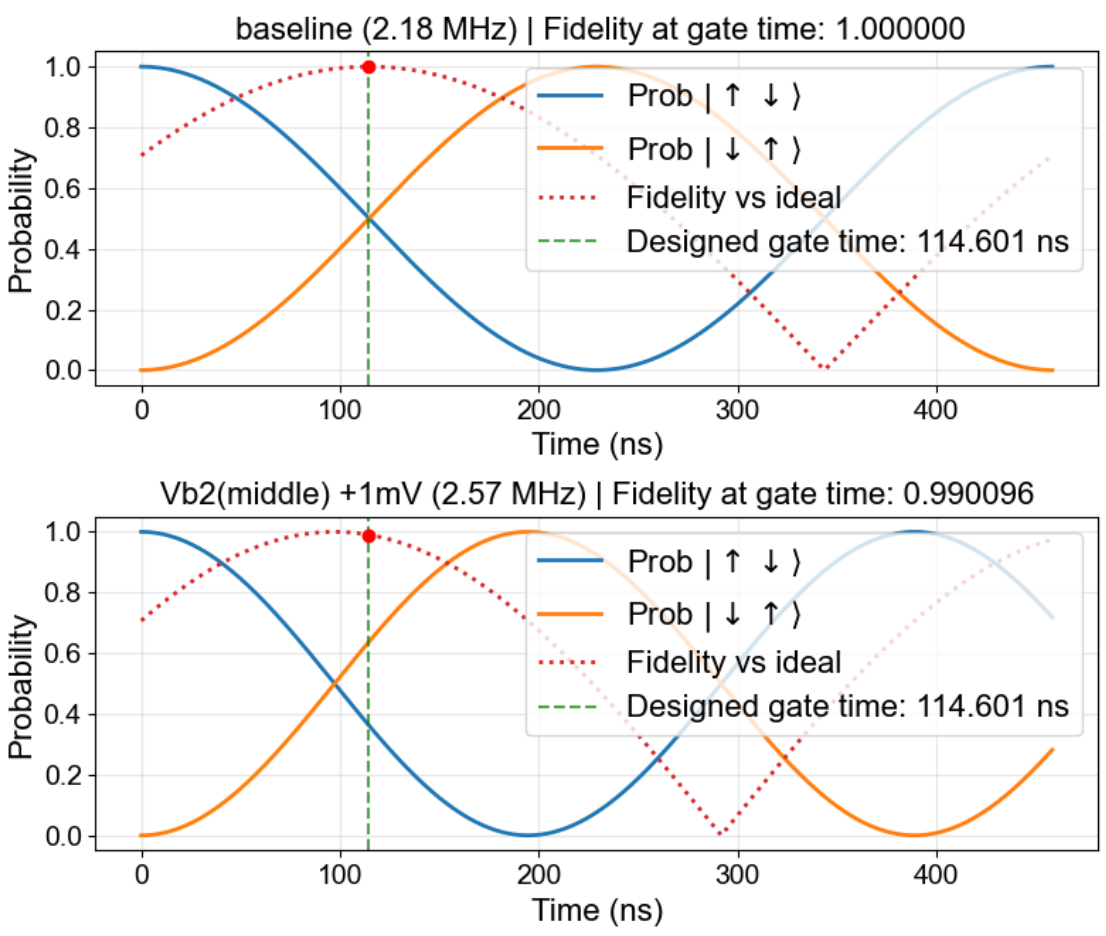

Fig. 9. Two-qubit state probability oscillations as a function of time under different $J$. Top: nominal condition $V_{B2}$ = 0.54 V with a nominal $\sqrt{SWAP}$ gate time = 114.601 ns. Bottom: An example of bias variation when $V_{B2}$ = 0.541 V. At the nominal $\sqrt{SWAP}$ gate time, its fidelity is only 0.99.

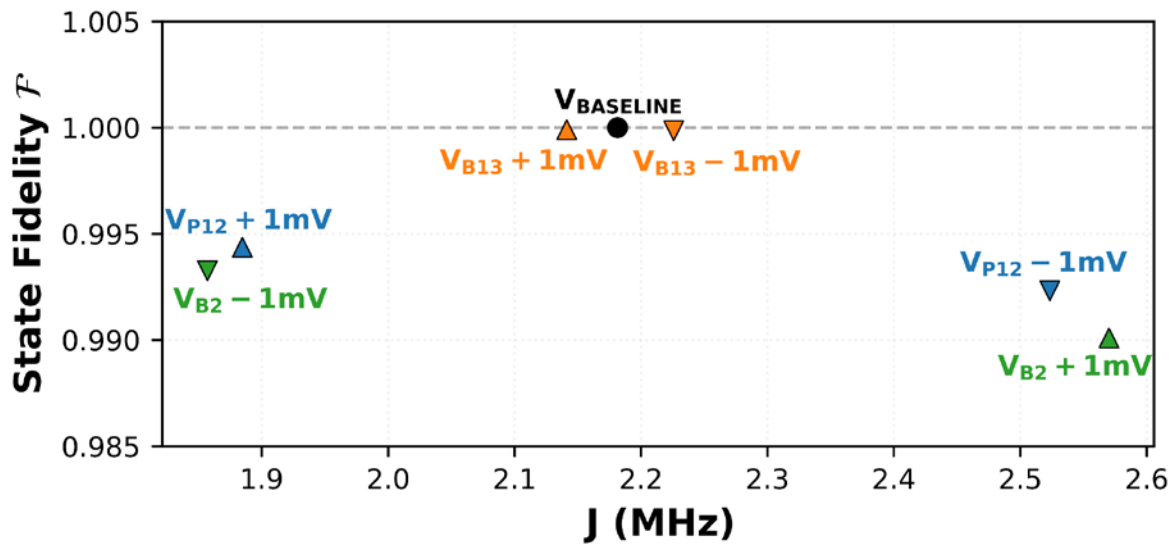

Fig. 10. Gate fidelity under ± mV bias fluctuations. The circle represents ideal operation; triangles and inverted triangles show fidelity changes due to the applied voltage perturbations.

To understand the gate-referred 1/f charge, $T_2 = \hbar/\sigma_v\sqrt{\sum_j (dJ/dV_j)^2}$ is calculated $\sigma_v \in [10^{-8}, 10^{-4}]V$ [16], where $\sigma_v$ is the chemical potential standard deviation. Each $dJ/dV_j$ term is shown in Table I.

TABLE I. SENSITIVITY OF J TO BIAS VOLTAGES USED TO EVALUATE THE COHERENCE TIME.

| | $V_{p1}$ | $V_{p2}$ | $V_{B1}$ | $V_{B2}$ | $V_{B3}$ | *Sum* |
|---|---|---|---|---|---|---|
| $dJ/dV$ (MHz) | -209 | -139 | -14 | 356.5 | 19.05 | 436.4 |

At $\sigma_v$ =$10^{-6}$ V, the coherence time was computed for both the full gate set and the central barrier gate. The results yield $T_2 \sim 364.7\ \mu s$ when summing over all gates, and $T_2 \sim 446.3\ \mu s$ when considering only the central barrier gate, which is consistent with the ratio reported in [16].

## IV. CONCLUSION

A nanosheet-based silicon spin-qubit DQD has been simulated using a 3D Schrödinger solver and a many-body exchange-interaction calculation. The nanosheet device provides a distinct electrostatic environment compared with FD-SOI, including a nearly constant potential profile along the nanosheet thickness for the same layout and silicon thickness. This requires adjusted biasing to achieve two-electron confinement. The simulations show that $J$ is highly sensitive to the plunger and middle-barrier gate biases because these biases set the effective inter-dot barrier. Geometry variations also matter, but their impact depends on the parameters: $t_{Si}$ variation mainly affects electron occupation, $L_B$ variation mainly affects exchange through tunnel coupling, and $L_P$ variation affects both occupation and exchange. Gate-level simulations based on the extracted $J$ values show that millivolt-level perturbations at $V_{P12}$ or $V_{B2}$ can reduce the fidelity below 99%. Overall, the results demonstrate that nanosheet technology is a promising platform for silicon spin qubits while identifying the electrostatic and process sensitivities that must be controlled for high-fidelity two-qubit gates.

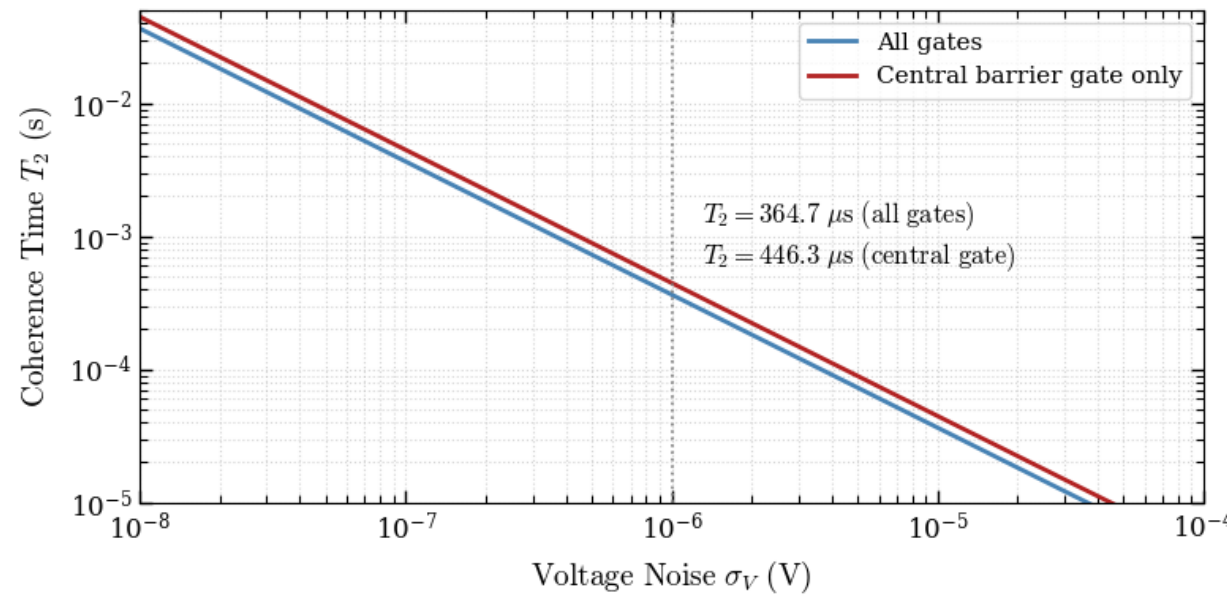

Fig. 11. Coherence time $T_2$ as a function of voltage-noise standard deviation.

## ACKNOWLEDGMENT

The authors are grateful for technical support and discussions with Mohammad Reza Mostaan of Nanoacademic Technologies Inc. This paper is the result of the work supported by the National Science Foundation under Grant No. 2125906 and AMDT Endowment Funding from the College of Engineering, San Jose State University.